\documentstyle[sprocl]{article}

\def\be{\begin{equation}}
\def\ee{\end{equation}}
\def\bea{\begin{eqnarray}}
\def\eea{\end{eqnarray}}

\begin{document}

\title{THE PROTON SPIN STRUCTURE IN A LIGHT-CONE
QUARK-SPECTATOR-DIQUARK MODEL\footnote{Invited talk presented
at the Diquarks 3rd Workshop, October 28-30, Villa Gualino, Torino,
Italy. To appear in the proceedings (World Scientific) edited by
M.~Anselmino and E.~Predazzi. hep-ph/9703425}}

\author{BO-QIANG MA}

\address{Institute of High Energy Physics, Academia Sinica\\
P.~O.~Box 918(4), Beijing 100039, China}


\maketitle
\abstracts{
It is shown that the proton spin problem raised by the
Ellis-Jaffe sum rule violation does not in conflict
with the SU(6) quark model, provided that the relativistic effect
from the quark
transversal motions, the flavor asymmetry between the $u$ and $d$
valence quarks, and the intrinsic quark-antiquark pairs
generated by the non-perturbative meson-baryon fluctuations
of the nucleon sea are taken into account.
The meson-baryon fluctuations of the nucleon sea provide 
a consistent picture for connecting the proton spin 
structure with a number of other anomalies
observed in the proton's structure, such as
the flavor asymmetry of the nucleon
sea implied by the violation of the Gottfried sum rule
and the discrepancy between two different measurements
of the strange quark distributions in the nucleon sea.  
}
  
\section{The Proton Spin Puzzle and the Wigner Rotation}

\subsection{The Violation of the Ellis-Jaffe Sum Rule 
and the Proton Spin
``Crisis''}

Parton sum rules and similar relations
played important roles in the establishment of the quark-parton
picture for nucleons in deep inelastic scattering.
Thus any violation of a parton sum rule may 
of essential importance to reveal possible new content concerning
our understanding of the underlying quark-gluon structure of hadrons.  

Form the SU(6) quark model one would expect
that the spin of the
proton is fully provided by the valence quark spins. Therefore
the observation of the Ellis-Jaffe sum rule violation received
attention by its implication that
the sum of the quark helicities is much smaller than the proton
helicity.
The EMC result of a much smaller integrated spin-dependent structure 
function data than that expected from the Ellis-Jaffe sum rule 
triggered the proton ``spin crisis'', i.e., the
intriguing question of how the spin of the proton is distributed among
its quark spin, gluon spin and orbital angular momentum. 
It is commonly taken for granted that the EMC result implies that
there must be some contribution due to gluon polarization or orbital
angular momentum to the proton spin. For example, in the gluonic 
and the strange sea explanations of the EJSR breaking, the proton
spin carried by the spin of quarks was estimated to be of about
$70\%$
in the former and of about $30\%$ in the latter\cite{Ans94}. 
It will be reported here based on previous works$^{2-5}$, 
however, that the 
the proton spin problem raised by the
Ellis-Jaffe sum rule violation does not in conflict
with the SU(6) quark model in which the spin of the proton, 
when viewed in its
rest reference frame, is provided by the vector sum of the
quark spins, 
provided that the relativistic effect
from the quark
transversal motions\cite{Ma91b,Bro94}, 
the flavor asymmetry between the $u$ and $d$
valence quarks\cite{Ma96}, and the intrinsic quark-antiquark pairs
generated by the non-perturbative meson-baryon fluctuations
of the nucleon sea\cite{Bro96} are taken into account.

\subsection{The Relativistic Effect due to Quark Transversal Motions}

As it is known, spin is essentially a relativistic notion associated with
the space-time symmetry of Poincar\'{e}.
The conventional 3-vector spin
{\bf s} of a moving particle with finite mass $m$ and 4-momentum $p_\mu$
can be defined by transforming its Pauli-Lub\'{a}nski 4-vector
$\omega_\mu=1/2 J^{\rho\sigma}P^\nu\epsilon_{\nu\rho\sigma\mu}$
to its rest frame via a non-rotation Lorentz boost $L(p)$ which
satisfies $L(p)p=(m,{\bf 0})$, by $(0,{\bf s})=L(p)\omega/m$.
Under an arbitrary Lorentz transformation, 
a particle state with spin ${\bf s}$
and 4-momentum $p_\mu$ will transform to the state with spin ${\bf s}'$ and
4-momentum $p'_\mu$,
\bea
{\bf s}'=R_\omega({\bf \Lambda},p){\bf s}, ~~~~ p'={\bf \Lambda}p,
\eea
where $R_\omega({\bf \Lambda},p)=L(p'){\bf \Lambda} L^{-1}(p)$
is a pure rotation known as Wigner rotation.
When a composite system is transformed from one frame to another one,
the spin of each constituent will undergo a Wigner rotation.
These spin rotations
are not necessarily the same since the constituents have different internal
motion. In consequence, the sum of the constituent's spin is not Lorentz
invariant. 

The key points for understanding the proton spin puzzle 
lie in the
facts that the vector sum of the constituent spins for a composite
system is not Lorentz invariant by taking into account the
relativistic effect of Wigner rotation, and that it is in the
infinite momentum frame the small EMC result was interpreted as
an indication that quarks carry a small amount of the total spin of
the proton. 
From the first fact we know that the vector spin structure of
hadrons could be quite different in different frames from
relativistic viewpoint. We thus can naturally understand the proton
``spin crisis'' because there is no need to require that the sum of 
the quark spins
be equal to the spin of the proton in the infinite
momentum frame, even if the vector sum of the quark spins equals to the
proton spin in the rest frame. 

The effect due to the Wigner rotation
can be best understood from the light-cone spin structure
of the pion. It has been shown\cite{Ma93,Hua94,Ma95}
that there are higher helicity
($\lambda_1+\lambda_2=\pm 1$) components in the light-cone
spin space wavefunction for the pion besides the
usual helicity ($\lambda_1+\lambda_2=0$) components\cite{Ma93,Hua94}. 
Therefore
the light-cone wavefunction for the lowest valence state
of pion can be expressed as
\begin{eqnarray}
|\psi^{\pi}_{q\overline{q}}>=
\psi(x,\vec{k}_{\perp},\uparrow,\downarrow)|\uparrow\downarrow>
+\psi(x,\vec{k}_{\perp},\downarrow,\uparrow)|\downarrow\uparrow>
\nonumber \\
+\psi(x,\vec{k}_{\perp},\uparrow,\uparrow)|\uparrow\uparrow>
+\psi(x,\vec{k}_{\perp},\downarrow,\downarrow)|\downarrow\downarrow>,
\label{eq:piwf}
\end{eqnarray} 
It is interesting to be noticed
that the light-cone wave function (\ref{eq:piwf}) 
is the 
correct pion spin wave function since it is an eigenstate
of the total spin operator $(\hat{S}^{F})^{2}$ 
in the light-cone formalism\cite{Ma95}. 

It is thus necessary to clarify what is meant by the quantity
$\Delta q$ defined by
$\Delta q\!\cdot\!S_{\mu}=<\!P,S|\bar{q}\gamma_{\mu}\gamma_{5}q|P,S\!>$, 
where
$S_{\mu}$ is the proton polarization vector. $\Delta q$ can be
calculated from $\Delta q=<\!P,S|\bar{q}\gamma^{+}\gamma_{5}q|P,S\!>$
since the instantaneous fermion lines do not contribute to the +
component. One can easily prove, by expressing the quark wave
functions in terms of light-cone Dirac spinors (i.e., the quark
spin states in the infinite momentum frame), that 
\begin{equation}
\Delta q=\int_{0}^{1}dx\:[q^{\uparrow}(x)-q^{\downarrow}(x)],
\end{equation}
where $q^{\uparrow}(x)$ and $q^{\downarrow}(x)$ are the probabilities
of finding, in the proton infinite momentum frame, a quark or
antiquark of flavor q with fraction x of the proton longitudinal
momentum and with polarization parallel or antiparallel to the proton
spin, respectively. However, if one expresses the quark wave
functions in terms of conventional instant form Dirac spinors
(i.e., the quark spin state in the proton rest frame), it can be
found, that 
\begin{equation}
\Delta q=\int\!d^{3}\vec{p}\:M_{q}\:[q^{\uparrow}(p)-q^{\downarrow}(p)]
=<\!M_{q}\!>\Delta q_{L},
\end{equation}
with 
\begin{equation}
M_{q}=[(p_{0}+p_{3}+m)^{2}-\vec{p}_{\perp}^{2}]/[2(p_{0}+p_{3})
(m+p_{0})]
\end{equation}
being the contribution from the relativistic effect due to the quark
transversal motions, $q^{\uparrow}(p)$ and $q^{\downarrow}(p)$ being the
probabilities of finding, in the proton rest frame, a quark or
antiquark of flavor $q$ with rest mass $m$ and momentum $p_{\mu}$ and
with spin parallel or antiparallel to the proton spin
respectively, and $\Delta q_{L}=\int\!d^{3}\vec{p}
[q^{\uparrow}(p)-q^{\downarrow}(p)]$ being the net spin vector sum of
quark flavor $q$ parallel to the proton spin in the rest frame. Thus
one sees that the quantity $\Delta q$ should be
interpreted as the
net spin polarization in the infinite momentum frame if one properly
considers the relativistic effect due to internal quark transversal
motions.

Since $<\!\!M_{q}\!\!>$, the average contribution from the relativistic
effect due to internal transversal motions of quark flavor $q$,
ranges from $0$ to $1$ (or more properly, it should be around
0.75 for light flavor quarks and approaches 1 for heavy flavor quarks), 
and $\Delta q_{L}$, the net spin vector polarization
of quark flavor $q$ parallel to the proton spin in the proton rest
frame, is related to the quantity $\Delta q$ by the relation 
$\Delta q_{L}=\Delta q/<\!\!M_{q}\!\!>$, we have sufficient freedom to make
the naive quark model spin sum rule, 
i.e., 
$\Delta u_{L}+\Delta d_{L}+\Delta s_{L}=1$, 
satisfied while still preserving 
the values of $\Delta u$,
$\Delta d$ and $\Delta s$ as parametrized from experimental
data in appropriate 
explanations. 
Thereby we can understand the ``spin crisis'' simply because the quantity
$\Delta\Sigma=\Delta u+\Delta d+\Delta s$ does not represent, in a strict
sense, the vector sum of the spin carried
by the quarks in the naive quark model.
It is possible that the value of
$\Delta\Sigma=\Delta u+\Delta d+\Delta s$ is small whereas the
spin sum rule
\begin{equation}
\Delta u_{L}+\Delta d_{L}+\Delta s_{L}=1 
\end{equation}
for the naive quark model still holds, 
though the realistic situation may be complicated.
 
\section{A Light-Cone Quark-Spectator-Diquark Model for Nucleons}

From the impulse approximation picture of deep inelastic scattering,
one can calculate the valence quark distributions  
in the quark-diquark model where the single valence quark
is the scattered parton and 
the non-interacting diquark serves to provide the quantum number
of the spectator\cite{Ma96}. 
The proton wave function in the quark-diquark model\cite{Pav76}
is written as 
\begin{equation}
\Psi _p^{\uparrow, \downarrow}(qD)
=sin\theta\, \varphi_{V}\, 
|qV>^{\uparrow, \downarrow} 
+ cos\theta\,
\varphi_{S}\, |qS>^{\uparrow, \downarrow},
\label{eq:pwf}
\end{equation}
with 
\begin{equation}
\begin{array}{clcr}

|qV>^{\uparrow, \downarrow}
=\pm \frac{1}{3}[V_0(ud)u^{\uparrow, \downarrow}-
\sqrt{2}V_{\pm 1}(ud)u^{\downarrow, \uparrow}
-\sqrt{2}V_0(uu)d^{\uparrow, \downarrow}
+2V_{\pm 1}(uu)d^{\downarrow, \uparrow}]; \\

|qS>^{\uparrow, \downarrow}
=S(ud)u^{\uparrow, \downarrow},
\label{eq:pwfVS}
\end{array}
\end{equation}
where $V_{s_z}(q_{1}q_{2})$ stays for a $q_1 q_2$ vector diquark 
Fock state with third spin 
component $s_z$, 
$S(ud)$ stays for a $ud$ scalar diquark Fock state,
$\varphi_{D}$
stays for the momentum space wave function of the 
quark-diquark with
$D$ representing 
the vector ($V$) or scalar ($S$) diquarks, 
and $\theta$ is a mixing angle that
breaks SU(6) symmetry at $\theta\neq \pi /4$. 
We choose the bulk SU(6) symmetry case $\theta=\pi /4$.
From Eq.~(\ref{eq:pwf}) we get the unpolarized quark distributions
\begin{equation}
\begin{array}{clcr}
u_{v}(x)=\frac{1}{2}a_S(x)+\frac{1}{6}a_V(x);\\
d_{v}(x)=\frac{1}{3}a_V(x),
\label{eq:ud}
\end{array}
\end{equation}
where $a_D(x)$ ($D=S$ or $V$) is normalized such
that $\int_0^1 d x a_D(x)=3$ and denotes the amplitude for the quark
$q$ is scattered while the spectator is in the diquark state $D$.
Therefore we can write, 
by assuming the isospin symmetry between the proton and the neutron,
the unpolarized structure functions for nucleons, 
\begin{equation}
\begin{array}{clcr}
F^p_2(x)=x s(x)+\frac{2}{9} x a_S(x)+\frac{1}{9} x a_V(x);\\
F^n_2(x)=x s(x)+\frac{1}{18} x a_S(x)+\frac{1}{6} x a_V(x),
\label{eq:Fpn}
\end{array}
\end{equation}
where $s(x)$ denotes the contribution from the sea. 

Exact SU(6) symmetry provides the relation $a_S(x)=a_V(x)$ 
which implies the valence flavor symmetry $u_{v}(x)=2 d_{v}(x)$. This
gives the prediction\cite{Kut71}  $F^n_2(x)/F^p_2(x)\geq 2/3$ for
all $x$ and is ruled out by the experimental
observation $F^n_2(x)/F^p_2(x) <  1/2$ for $x \to 1$.
It has been a well established fact that the valence flavor
symmetry $u_{v}(x)=2 d_{v}(x)$ does not hold and the explicit
$u_{v}(x)$ and $d_{v}(x)$ can be parameterized from  
the combined
experimental data from deep inelastic 
scatterings of electron (muon) and neutrino (anti-neutrino) 
on the proton and the neutron {\it et al.}.  
In this sense, any theoretical calculation of quark distributions 
should reflect the flavor asymmetry between 
the valence $u$ and $d$ quarks in a reasonable picture.
It has been shown\cite{Ma96} that the mass
difference between the scalar
and vector spectators can reproduce the up and down valence
quark asymmetry that 
accounts for the observed ratio $F_2^{n}(x)/F_2^{p}(x)$ at large $x$.

The amplitude for the quark
$q$ is scattered while the spectator in the spin state $D$
can be written as
\begin{equation}
a_D(x)
\propto \int [{\rm d}^2 {\bf k}_{\perp}] |\varphi_{D}(x,{\bf k}_{\perp})|^2.
\end{equation}
We adopt the %
Brodsky-Huang-Lepage prescription 
for the
light-cone momentum space wave function\cite{BHL} of the 
quark-spectator 
\begin{equation}
\varphi_{D}(x,{\bf k}_{\perp})
=A_{D} exp\{-\frac{1}{8\beta^2_{D}}[\frac{m^2_q+{\bf k}^2_{\perp}}{x}
+\frac{m^2_D+{\bf k}^2_{\perp}}{1-x}]\},
\label{eq:BHL}
\end{equation}
where ${\bf k}_{\perp}$ is the internal quark transversal momentum,
$m_q$ and $m_D$ are the masses for the quark $q$ and spectator $D$,
and $\beta_D$ is the harmonic oscillator scale parameter.
The values of the parameters $\beta_D$, $m_q$ and $m_D$ 
can be adjusted by fitting the hadron properties
such as the electromagnetic form factors, 
the mean charge radiuses, and the
weak decay constants {\it et al.} in the relativistic light-cone
quark model. 
We simply adopt $m_q=330$~MeV and
$\beta_D=330$~MeV. 
The masses of the scalar and vector spectators should
be different taking into account the spin force from color magnetism
or alternatively from instantons\cite{Web92}. 
We choose, e.g., $m_S=600$ MeV and $m_V=900$ MeV as estimated
to explain the N-$\Delta$ mass difference. The mass difference
between the scalar and vector spectators
causes difference between $a_S(x)$ and $a_V(x)$ and 
thus the flavor asymmetry between the valence quark distribution
functions $u_v(x)$ and $d_v(x)$.
The calculated
results\cite{Ma96} are in reasonable agreement 
with the experimental data and this
supports the quark-spectator picture of deep inelastic scattering
in which the difference between the scalar and vector
spectators is important to reproduce the explicit
SU(6) symmetry breaking while the bulk SU(6) symmetry of the 
quark model still holds. 

For the polarized quark distributions,
we take into account the contribution from the
Wigner rotation\cite{Ma91b,Bro94}.
In the light-cone or quark-parton descriptions,
$\Delta q (x)=q^{\uparrow}(x)-q^{\downarrow}(x)$,
where $q^{\uparrow}(x)$ and $q^{\downarrow}(x)$ are the probability
of finding a quark or antiquark with longitudinal momentum
fraction $x$ and polarization parallel or antiparallel
to the proton helicity in the infinite momentum frame.
However, in the proton rest frame, one finds,
\begin{equation}
\Delta q (x)
=\int [{\rm d}^2{\bf k}_{\perp}] W_D(x,{\bf k}_{\perp}) 
[q_{s_z=\frac{1}{2}}
(x,{\bf k}_{\perp})-q_{s_z=-\frac{1}{2}}(x,{\bf k}_{\perp})],
\end{equation}  
with
\begin{equation}
W_D(x,{\bf k}_{\perp})=\frac{(k^+ +m)^2-{\bf k}^2_{\perp}}
{(k^+ +m)^2+{\bf k}^2_{\perp}} 
\end{equation}
being the contribution from the relativistic effect due to
the quark transversal motions, 
$q_{s_z=\frac{1}{2}}(x,{\bf k}_{\perp})$
and $q_{s_z=-\frac{1}{2}}(x,{\bf k}_{\perp})$ being the probability
of finding a quark and antiquark with rest mass $m$
and with spin parallel and anti-parallel to the rest proton
spin, and $k^+=x {\cal M}$ where 
${\cal M}=\frac{m^2_q+{\bf k}^2_{\perp}}{x}
+\frac{m^2_D+{\bf k}^2_{\perp}}{1-x}$.
The Wigner rotation factor $W_D(x,{\bf k}_{\perp})$ ranges
from 0 to 1; thus $\Delta q$ measured 
in polarized deep inelastic scattering cannot be  
identified
with the spin carried by each quark flavor in the proton
rest frame. 

From Eq.~(\ref{eq:pwf}) we get the spin distribution probabilities
in the quark-diquark model
\begin{equation}
\begin{array}{clcr}
u_V^{\uparrow}=\frac{1}{18};\;\;
u_V^{\downarrow}=\frac{2}{18};\;\;
d_V^{\uparrow}=\frac{2}{18};\;\;
d_V^{\downarrow}=\frac{4}{18};\\
u_S^{\uparrow}=\frac{1}{2};\;\;
u_S^{\downarrow}=0;\;\;
d_S^{\uparrow}=0;\;\;
d_S^{\downarrow}=0.\\
\label{eq:udVS}
\end{array}
\end{equation}
Taking into account the Wigner rotation,
we can write the quark helicity distributions
for the $u$ and $d$ quarks 
\begin{equation}
\begin{array}{clcr}
\Delta u_{v}(x)=u_{v}^{\uparrow}(x)-u_{v}^{\downarrow}(x)=
-\frac{1}{18}a_V(x)W_V(x)+\frac{1}{2}a_S(x)W_S(x);\\
\Delta d_{v}(x)=d_{v}^{\uparrow}(x)-d_{v}^{\downarrow}(x)
=-\frac{1}{9}a_V(x)W_V(x),
\label{eq:sfdud}
\end{array}
\end{equation}
where $W_D(x)$ is the correction factor due the Wigner rotation.
From Eq.~(\ref{eq:ud}) one gets 
\begin{equation}
\begin{array}{clcr}
a_S(x)=2u_v(x)-d_v(x);\\
a_V(x)=3d_v(x).
\label{eq:qVS}
\end{array}
\end{equation}
Combining Eqs.~(\ref{eq:sfdud}) and (\ref{eq:qVS}) we have
\begin{equation}
\begin{array}{clcr}
\Delta u_{v}(x)
    =[u_v(x)-\frac{1}{2}d_v(x)]W_S(x)-\frac{1}{6}d_v(x)W_V(x);\\
\Delta d_{v}(x)=-\frac{1}{3}d_v(x)W_V(x).
\label{eq:dud}
\end{array}
\end{equation}
Thus we arrive at simple relations between the polarized
and unpolarized quark distributions for the valence $u$ and $d$
quarks. We can calculate the quark helicity distributions
$\Delta u_{v}(x)$ and $\Delta d_{v}(x)$ from the unpolarized
quark distributions $u_{v}(x)$ and $d_{v}(x)$ by relations
(\ref{eq:dud}),
once the detailed $x$-dependent Wigner rotation
factor $W_D(x)$ is known. On the other hand, we can also
use relations (\ref{eq:dud}) to study 
$W_S(x)$ and $W_V(x)$, once there are good quark distributions 
$u_{v}(x)$, $d_{v}(x)$,
$\Delta u_{v}(x)$, and $\Delta d_{v}(x)$ from experiments.
From another point of view, the relations (\ref{eq:dud})
can be considered as the results of the conventional
SU(6) quark model 
by explicitly taking into account the Wigner rotation effect
and the flavor asymmetry introduced by the
mass difference between the scalar and vector
spectators, 
thus any evidence for the 
invalidity of Eq.~(\ref{eq:dud}) will be useful to reveal 
new physics beyond the SU(6) quark model.

We calculated the $x$-dependent Wigner rotation factor
$W_D(x)$ in the light-cone SU(6) quark-spectator  
model\cite{Ma96} and noticed slight asymmetry between $W_S(x)$ 
and $W_V(x)$.
Considering only the valence quark contributions, 
we  can write the 
spin-dependent structure functions $g_1^p(x)$ and $g_1^n(x)$
for the proton and the neutron by
\begin{equation}
\begin{array}{clcr}
g_1^p(x)=\frac{1}{2}[\frac{4}{9}\Delta u_v(x)
+\frac{1}{9}\Delta d_v(x)]
=\frac{1}{18}[(4 u_v(x)-2 d_v(x))W_S(x)-d_v(x)W_V(x)];\\
g_1^n(x)=\frac{1}{2}[\frac{1}{9}\Delta u_v(x)
+\frac{4}{9}\Delta d_v(x)]
=\frac{1}{36}[(2 u_v(x)-d_v(x))W_S(x)-3 d_v(x)W_V(x)].
\label{eq:g1pn}
\end{array}
\end{equation}
We found\cite{Ma96} that the calculated $A_1^N$
with Wigner rotation 
are in agreement  
with the experimental data, at least for $x \geq 0.1$.
The large asymmetry between $W_S(x)$ and $W_V(x)$
has consequence for a better fit of the data.

As we consider only the valence quark contributions
to $g_1^p(x)$ and $g_1^n(x)$, we should not expect to fit
the Ellis-Jaffe sum data from experiments. This leaves
room for additional contributions from sea quarks
or other sources.
We point out, however, it is possible to reproduce
the observed Ellis-Jaffe sums $\Gamma_1^p=\int_0^1 g_1^p(x){\rm d} x$
and $\Gamma_1^n=\int_0^1 g_1^n(x){\rm d} x$ 
within the light-cone SU(6) quark-spectator model
by introducing a large asymmetry between the Wigner rotation
factors $W_S$ and $W_V$ for the scalar and vector spectators.
For example, we need $<W_S>=0.56$ and $<W_V>=0.92$
to produce $\Gamma_1^p=0.136$ and $\Gamma_1^n=-0.03$
as observed in experiments. 
This can be achieved by adopting a large difference between 
$\beta_S$ and $\beta_V$ which should be adjusted by 
fitting other nucleon properties in the model\cite{Web92}.
The calculated
$A_1^p(x)$, $A_1^n(x)$, and $A_1^d(x)$ are
in good agreement with the data\cite{Ma96}. 
This may suggest that the explicit
SU(6) asymmetry could be also used to explain the EJSR
violation (or partially) 
within a bulk SU(6) symmetry scheme of the quark model, or
we take this as a hint for other SU(6) breaking source in additional to
the SU(6) quark model.

We showed in the above that the  $u$ and $d$ asymmetry
in the lowest valence component of the nucleon and the Wigner
rotation effect due to the internal quark transversal motions
are important for
re-producing 
the observed ratio
$F_2^n/F_2^p$ and the polarization asymmetries
$A_1^N$ for the proton, neutron, and deuteron.
For a better understanding of 
the origin of polarized sea quarks implied by
the violation of the  Ellis-Jaffe sum rule, 
we still need to
consider the higher Fock states implied by the non-perturbative
meson-baryon fluctuations\cite{Bro96}.
In the light-cone meson-baryon fluctuation model, the net $d$ quark
helicity of the intrinsic $q \bar q$ fluctuation is negative,
whereas the net $\bar d$ antiquark helicity is zero. Therefore the
quark/antiquark asymmetry of the $d \bar d$ pairs should be 
apparent in the $d$ quark and antiquark helicity distributions.
There are now explicit measurements of the helicity distributions
for the individual $u$ and $d$ valence and sea quarks by SMC\cite{SMC96}.  
The helicity distributions
for the $u$ and $d$ antiquarks are consistent with zero in agreement
with the results of the light-cone meson-baryon fluctuation model of
intrinsic $q \bar q$ pairs.
The calculated quark helicity distributions 
$\Delta u_{v}(x)$ and $\Delta d_{v}(x)$ have been compared\cite{Ma96} 
with the
recent SMC data.
The data are still not precise enough for making detailed comparison,
but the agreement with $\Delta u_{v}(x)$ seems to be good.
It seems that the agreement with $\Delta d_{v}(x)$ is poor
and there is somewhat evidence for additional source of negative
helicity contribution to the valence $d$ quark beyond the 
conventional quark model. 
This again
supports the light-cone meson-baryon fluctuation model in which the
helicity distribution of the intrinsic $d$ sea quarks $\Delta
d_s(x)$ is negative.

The standard SU(6) quark model gives the constraints $|\Delta u_v|
\leq \frac{4}{3}$ and $|\Delta d_v| \leq \frac{1}{3}$. 
A global fit\cite{Ell95b} 
of polarized deep inelastic scattering data together
with constraints from nucleon and hyperon decay and the included
higher-order perturbative QCD corrections leads to values for
different quark helicity contributions in the proton: $\Delta
u=0.83\pm 0.03, \;\; \Delta d=-0.43 \pm 0.03, \;\; \Delta s=-0.10
\pm 0.03.$ In the light-cone meson-baryon fluctuation model, the
antiquark helicity contributions are zero. We thus  can consider the
empirical values  as the helicity contributions $\Delta q=\Delta
q_v+\Delta q_s$ from both the valence $q_v$ and sea $q_s$ quarks.
Thus the empirical result  $|\Delta d| > \frac{1}{3}$ strongly
implies an additional negative contribution $\Delta d_s$ in the
nucleon sea.
 
\section{The Meson-Baryon Fluctuations of the Nucleon
Sea}
 
The light-cone meson-baryon fluctuation model of
intrinsic $q \bar q$ pairs\cite{Bro96} leads to a consistent picture
for understanding a number of empirical anomalies related to the
composition of the nucleons in terms of their non-valence sea quarks.
We have studied the sea quark/antiquark asymmetries in
the nucleon wavefunction which are generated by a light-cone model
of energetically-favored meson-baryon fluctuations.  The model
predicts striking quark/antiquark asymmetries in the momentum and
helicity distributions for the down and strange contributions to the
proton structure function: the intrinsic $d$ and $s$ quarks in the
proton sea are negatively polarized, whereas the intrinsic $\bar d$
and $\bar s$ antiquarks give zero contributions to the proton spin.
Such a picture is supported by experimental phenomena related to the
proton spin problem: the recent SMC measurement of helicity
distributions for the individual up and down valence quarks and sea
antiquarks, the global fit of different quark helicity contributions
from experimental data, and the negative strange quark helicity from
the $\Lambda$ polarization in $\bar{\nu} {\rm N}$ experiments by the
WA59 Collaboration. The light-cone meson-baryon fluctuation model
also suggests a structured momentum distribution asymmetry for
strange quarks and antiquarks which is related to an outstanding
conflict between two different measures of strange quark sea in the
nucleon\cite{Ma96b}. 
The model predicts an excess of intrinsic $d \bar d$ pairs
over $u \bar u$ pairs, as supported by the Gottfried sum rule
violation\cite{GSV}. We also predict that the intrinsic charm and anticharm
helicity and momentum distributions are not identical.
 
The intrinsic sea model thus gives a clear picture of quark flavor
and helicity distributions supported qualitatively by a number of
experimental phenomena.  It seems to be an important physical source
for the problems of the Gottfried sum rule violation, the
Ellis-Jaffe sum rule violation, and the conflict between two
different measures of strange quark distributions.

\section{Conclusion}

The ``intrinsic" $q \bar q$ pairs generated by the
non-perturbative 
meson-baryon fluctuations in the nucleon
sea are important source for a consistent picture of a
number of empirical anomalies.
The violations of the Gottfried sum rule and the
Ellis-Jaffe sum rule are closely connected by the
meson-baryon fluctuations in the nucleon, in contrary
to most studies in which they are not related.
The violation of the Ellis-Jaffe sum rule does not in
conflict with the SU(6) quark model, provided that the relativistic
effect from the Wigner rotation, the flavor asymmetry generated 
by the spin-spin interactions of the valence quarks, and
the intrinsic $q \bar q$ pairs generated by 
non-perturbative meson-baryon fluctuations are taken
into account.
One important prediction of the model 
is the significant
quark/antiquark asymmetries in the momentum and helicity distributions
for the quarks and antiquarks of the nucleon sea, contrary to
intuition based on perturbative gluon-splitting processes.   
It is important to test the meson-baryon fluctuation model 
by it's predictions in various physical processes, such as  
the forward energetic
neutron events at HERA and the asymmetric quark/antiquark
hadronization in $e^+e^-$ annihilation\cite{Bro97}.

\section*{Acknowledgments}
I am very grateful to Professors Stan J.~Brodsky, Tao Huang,
and Qi-Ren Zhang for their valuable contribution, guidance and 
encouragement during the profitable and enjoyable
collaborations from which 
the contents in this talk were developed.

\end{document}